\documentclass[12pt,preprint]{aastex}

\shorttitle{Improved version of the IIM}
\shortauthors{Simonneau et al.}

\begin{document}
\title{An improved version of the Implicit Integral Method to
solving radiative transfer problems}

\author{E. Simonneau}
\affil{Instituto de Astrof\'isica de Canarias, IAC-ULL, V\'ia Lactea s/n, 
E-38205 La Laguna,Tenerife, Spain and 
Institut d'Astrophysique de Paris, CNRS-UPMC, 98 bis Boulevard
Arago, F-75014 Paris, France}

\author{O. Cardona}
\affil{Instituto Nacional de Astrof\'isica, \'Optica y Electr\'onica,
L. E. Erro 1, M-72840 Tonantzintla, Puebla, Mexico}
\email{ocardona@inaoep.mx}

\and

\author{L. Crivellari}
\affil{Instituto de Astrof\'isica de Canarias, V\'ia Lactea s/n, 
E-38205 La Laguna,Tenerife, Spain and Depto. Astrof\'isica, Universidad de
La Laguna , E-38206 La Laguna, Tenerife, Spain} 
\email{luc@iac.es}

\begin{abstract}
Radiative transfer (RT) problems in which the source function includes a 
scattering-like integral are typical two-points boundary problems. Their 
solution via differential equations implies to make hypotheses on the
solution itself, namely the specific intensity $I(\tau;\bf{n})$ of the
radiation field.  On the contrary, integral methods require to make hypotheses
on the source function $S(\tau)$. It looks
of course more reasonable to make hypotheses on the latter
because one can expect that the run of $S(\tau)$ with depth be smoother than 
that of $I(\tau;\bf{n})$.

In previous works we assumed a piece-wise 
parabolic  approximation for the source function, which warrants the continuity
of $S(\tau)$ and its first derivative at each depth point.
Here we impose the continuity of the second derivative $S^{\prime\prime}(\tau)$.
In other words, we adopt a cubic spline representation to the source function,
which highly stabilize the numerical processes. 

\end{abstract}


\section{Introduction}

Some years ago we proposed a new algorithm, the Implicit Integral
Method (IIM), to solving those radiative transfer  problems in which the
specific source functions (one for each frequency and direction pair)
depend linearly on the radiation field via a single
quantity independent of both frequency and direction. In the paradigm instance 
of radiative transfer through an ideal medium formed by atoms with only two
energy levels (Two-Level Atom model), this quantity is the integral over
frequencies of the mean specific intensity of the radiation field, weighted
with the spectral profile. (See Simonneau and Crivellari, 1993, hereinafter
Paper I.)

Because it is independent of both frequency and direction, such a quantity
constitutes a single scalar coupling for all the specific RT equations, and
can be chosen in a natural way as the protagonist variable for the numerical
solution of the RT problem. This choice is the distinctive and essential
feature of our IIM: to work with a quantity which is independent of both
frequency and direction brings about that the method does not require to store
and invert huge matrices like in the customary numerical algorithms employed
in RT problems. 
We have already remarked in Paper I that our algorithm is a mere 
phenomenological representation of the actual physical process. Because of 
that and due to the lack of a matricial structure, the advantages of the IIM
in terms of reliability, accuracy and robustness sould be self-evident, as
well as the conspicous saving of both computational time and memory storage
it makes possible.

The aforesaid advantages suggested us the possibility to employ the IIM also
in the computation of stellar atmospheres models, where we must solve
many (some hundreds) RT equations, one for each frequency. The source function
of each specific RT equation is here the weighted mean of a 
term that includes the mean specific intensity of the radiation field through a
scattering-like
integral with a thermal contribution given by the Planck function $B_{\nu}(T)$.
The paradigm problem of the self-consistent temperature correction when
computing stellar atmosphere models was considered in Crivellari and Simonneau 
(1994).

First of all, we must recognize that the geometrical
structure of the system, that is the sequence of the discrete atmospheric
layers, must be necessarily the same for all the frequencies. But we must also
recognize that for any given frequency some layers do not contribute to the
formation of the spectrum. They do not take an effective part in the radiative
transfer process because either they are exceedingly transparent ({\it{i.e.}}
$\exp \lbrace - \Delta\tau_{\nu}\rbrace \simeq 1$) 
or they correspond to optically very deep regions ({{\it{i.e.}} 
$\exp \lbrace - \Delta\tau_{\nu}\rbrace \simeq 0$). The layers
intermediate between the above two groups constitute the specific spectral
formation region.
However all the layers of the structure must be taken into account in the
numerical algorithm, irrespectively of the frequency considered.
Yet, due to the dramatic difference among the values of the opacity with 
frequency,
different spectral intervals form in very different geometrical regions.
That compels us to divide the atmosphere into very many
layers in order to cover properly all the spectral formation intervals.
On the other hand, it is matter of the run with depth of the data that are 
common to radiative transfer at all the frequencies. As an exemple, given a 
temperature distribution on the discrete atmospheric layers,
the variation with depth of the
numerical values of the Planck function $B_{\nu}(T)$, {\it{i.e.}} 
the monochromatic thermal sources,
may vary enormously frequency by frequency. 
For instance, in the case of a solar-like star, from the bottom to 
the top of the atmosphere $B_{\nu}(T)$ varies by a factor of
the order of $10^{3}$ for frequencies in the visible part of the spectrum,
while this factor can be of the order of $10^{14}$ for frequencies in the
range of Lyman{} ${\alpha}$. Therefore a set of depth 
points suitable for a good description of the mathematical behaviour of the
source function  at some frequencies cannot be adequate at other
frequencies. Again very many common discrete depth points are necessary
in order to provide a proper distribution of the data for the adequate
treatment of each monochromatic RT equation.

The foregoing requirements make it impossible in the practice to replace
derivatives by finite differences, as in the outermost layers the optical
thickness is almost zero for many frequencies. The use of integral methods may 
seem to be the only advisable way out, but the very large number of discrete
optical depth points, necessary to warrant the proper treatment of the RT
process at all the frequencies, does not advise to employ global integral
methods, too.

We can get rid of the difficulties brought about by the introduction of very
many layers on the one hand by employing our IIM,
which allows us to take into consideration as many geometrical depths as 
necessary because, as already said, it does not require the storage and 
inversion of huge matrices. Moreover, on the other hand, we can introduce a
better mathematical representation of each monochromatic source function
$S_{\nu}(\tau_{\nu})$ in order to account for the possible
rapid variation of both the
branching parameter $\varepsilon_{\nu}$ (see eq. [2] later)
and $B_{\nu}(T)$ with respect to each
specific optical depth $\tau_{\nu}$. In such a way we can optimize the treatment
of all the individual frequencies.

In the original formulation of the IIM (see the above references) we considered
models that comprised 150-200 discrete layers between the surface and the
bottom of the atmosphere. Inside each of them we approximated each specific
source function $S_{\nu}(\tau_{\nu})$ by an  arc of parabola and imposed the
continuity of $S_{\nu}(\tau_{\nu})$ and 
$S^{\prime}_{\nu}(\tau_{\nu})$ at all the {\it{NL}} dividing
points. This piece-wise parabolic approximation yielded excellent results in 
many cases (see the above references). However under extreme conditions, for 
instance in the case of sudden variations of the thermal sources
({\it{e.g.}} at Lyman $\alpha$ frequencies in cool stars), such an 
approximation 
can introduce numerical instabilities that spoil the computation of the model.

To impose also the continuity of the second derivative of 
$S_{\nu}(\tau_{\nu})$ at
all the {\it{NL}} dividing points can remove the foregoing instabilities. 
Consequently we propose here a cubic spline model for each specific source
function. In some way this model constitutes a regularization of the process
to computing the values of the source functions.
The formalism of the cubic spline approximation (namely a two-point boundary
value problem developed to interpolate among the {\it{NL}}
explicitly known values of a given function) can be employed in the present 
case although the {\it{NL}} values of $S_{\nu}(\tau_{\nu})$ are yet unknown.

To employ the cubic spline approach in order to describe the behaviour of the
source function in typical RT problems, where a scattering term appears in the
source function, is the best (may be the unique) correct choice for both 
theoretical
and numerical reasons. A theoretical reason is brought about by the non-local
nature of the problem: the specific intensities and consequently the source 
function at a given depth point depend via the RT process
on the values of the source function at
all the other points of the system. Thus the numerical values of the source
function must be computed simultaneously at all the depth points.
Therefore such a non-local character of the physical problem must be represented
by means of a non-local mathematical structure.
Also the derivatives of the source function at
any depth point must be formulated as a linear relation including the implicit
values of the source function at all the depth points, not only as a linear 
relation of the implicit values of the source function at each triad of
consecutive depth points. 

The pratical reason is for the sake of the stability of the computational
algorithm. The cubic spline model minimizes the strain energy integral, that is
the integral of the squared values 
of the values of the second derivative of the protagonist function, namely of
the variation of its curvature - {\it{i.e.}} the oscillations. 
(See, {\it{e.g.}},
Rivlin, 1981.) That is, the use of the cubic spline approximation to the source 
function minimizes the risk of destabilizing oscillations.
The cubic spline representation constitutes by itself another two-points 
boundary problem:
we must know the values of $S(\tau)$ at all the depth points in order to
compute $S^{\prime}(\tau)$ and $S^{\prime\prime}(\tau)$ 
at each depth point. Nevertheless this
apparent drawback turns out to be on the contrary an advantage, because we
can carry on simultaneously both two-points boundary problems: the 
RT problem and the cubic spline interpolation.

From the algorithmical stand point, the kernel of the original IIM is a 
forward-elimination scheme that links the so far
unknown values of the source function  at each pair of consecutive
optical depth points $(\tau_{L}, \tau_{L+1})$ by mean of a linear relation with
known coefficients. The latter are determined by taking into account
the RT equations that describe layer by layer the propagation of both
the downgoing and the upgoing specific intensities. Now we realized that, by
using the cubic spline formalism the same 
forward-elimination scheme can also be employed to link 
the unknown values of the second derivatives of the source functions, again
by means of a linear relation.

Once attained the deepest optical depth point $\tau_{NL}$ at the end of the
forward-elimination, we can impose the
bottom boundary condition (eq. [5] later on) to both the RT process and the 
cubic spline chain; in other words we can close the linear relation
between $S_{\nu}(\tau_{NL-1})$ and $S_{\nu}(\tau_{NL})$ on the one hand, 
between $S^{\prime\prime}_{\nu}(\tau_{NL-1})$ and 
$S^{\prime\prime}_{\nu}(\tau_{NL})$
on the other. This allows us to recover the numerical 
values of the source functions and their second derivatives at the bottom,
as well as those of the set of the
incident upgoing specific intensities 
$\lbrace I^{+}(\tau_{NL},\mu_{J}), J = 1,ND\rbrace$.
Then, in  a succesive back-substitution scheme, we are in a position to compute
at each depth point the numerical values of the source functions and their 
second derivatives  by using the above linear relations,
whose coefficients have been stored during the previous forward-elimination.

Thanks to that we have at hand a unique algorithm to solve each specific RT 
problem under the
imposed constraint that the specific source functions as well as their first and
second derivatives be continuous at all the {\it{NL}} points of the grid chosen
for the geometrical representation of the stellar atmosphere. In such a way we 
can get rid of the instabilities that may arise in the case of extreme 
variations of the source functions without paying any extra computational cost.

\section{The mathematical background}

For the sake of an easier presentation of the new more precise version of the
IIM announced in Section 1, we will consider the simplest instance
that yet contains all the difficulties intrinsic to RT astrophysical problems,
namely the transport of monochromatic radiation through a plane-parallel medium
in which matter particles can scatter, absorb and emit photons. In the previous
works above quoted
the original formulation was applied to much more general instances.
The version presented here can be easily applied to such cases.

Following the customary notation, the RT
equations that describe the evolution of  the upgoing intensities 
$I^{+}(\tau,\mu)$ and the downgoing intensities $I^{-}(\tau,\mu)$ are
 
\begin{equation}
\pm \mu\ \frac{d }{d\tau} I^{\pm}(\tau,\mu)\\ = \ I^{\pm}(\tau,\mu)\ -\ S(\tau)
\ ,
\end{equation}

\parindent=0em
where $\tau$ denotes the optical depth and $\mu$ is the cosine of the angle formed
by the direction of propagation with the perpendicular to the plane-parallel 
layers ($\mu \equiv \cos \theta, 0 \leq \mu \leq 1)$ 

\parindent=3em
The source function is a weighted  mean between the thermal source
$B({\tau})$ and the mean intensity $J(\tau)$, namely

\begin{equation}
S(\tau) \ = \ \varepsilon B(\tau) \ + \ (1 - \varepsilon) J(\tau)\ .
\end{equation}

The branching parameter $\varepsilon \ = \ \varepsilon(\tau)$ is the ratio of
the absorption coefficient to the total opacity ({\it{i.e.}} the sum of the
absorption and the scattering coefficient). The latter defines the 
scale of the optical depth $\tau$; ($1 - \varepsilon$) is customarily called 
the albedo. In terms of the upgoing and the downgoing intensities the mean
intensity is given by

\begin{equation}
J(\tau) \ = \ \int_{0}^{1} [I^{+} (\tau,\mu) \ + \ I^{-} (\tau,\mu)]\ d \mu\ .
\end{equation}

\parindent=0em
The integral in eq. (3) is representative of any scattering integral, which may
be different for the application of the IIM to different instances.  

\parindent=3em   
In the discrete ordinates approximation the integral in eq. (3) is replaced by 
the sum of the intensities corresponding to a finite number of $ND$ directions.
Then 

\begin{equation}
J(\tau) \ \approx \sum_{J = 1}^{ND} {\bf{w}}_{J} [I^{+}(\tau,\mu_{J}) \ + 
\ I^{-}(\tau,\mu_{J})]\ .
\end{equation}

For most RT problems in plane-parallel geometry (at least for stellar atmosphere
models computations) a five-points Gauss division of the interval 
$0\ \leq\ \mu\ \leq\ 1$ is more than enough. The ${\bf{w}}_{J}$'s are the
corresponding integration weights.
 
The numerical solution requires the discretization of the optical depth variable
$\tau$, too. The stellar atmosphere must be sliced into a set of 
${NL}$ plane-parallel horizontal layers, divided by the set of 
$NL + 1$ optical depths points 
$\lbrace \tau_{0},\tau_{1},\tau_{2}...\tau_{NL}\rbrace$. The
value $\tau_{0} = 0$ corresponds to the surface and $\tau_{NL}$ to the 
bottom of the atmosphere. 
The computation of a fairly good model require that ${NL}$ be of the
order of two hundred.

The values of the incident intensities onto the top surface, 
{\it{i.e.}} the downgoing intensities  
$I^{-}(\tau_{0},\mu_{J})$,
and those of the incident intensities onto the bottom surface,
{\it{i.e.}} the upgoing intensities  
 $I^{+}(\tau_{NL},\mu_{J})$, must be known; they are data of the RT problem. 
In the case of a stellar atmosphere $I^{-}(\tau_{0},\mu)$
is usually
assumed to be zero, that is  there is not radiation incident onto the
stellar surface. 
We will show later that the method can equally work also under more general
conditions.
For the upgoing intensities at the bottom of the atmosphere we
can assume that the diffusion approximation holds valid, that is

\begin{equation}
I^{+}(\tau _{NL},\mu _{J})\ =\ S(\tau
_{NL})\ +\ S^{\prime}(\tau _{NL})\ \mu_{J} \ + \ S^{\prime\prime}(\tau_{NL})\ 
\mu^{2}_{J}\ +\ S^{\prime\prime\prime}(\tau_{NL})\ \mu ^{3}_{J}\ ,
\end{equation}

\parindent=0em
which is brought about by the cubic polynomial behaviour of $S(\tau)$ at
depths immediately greater than $\tau_{NL}$.
These two families of boundary conditions are sufficient to ensure that the RT
problem is self-consistent.

\parindent=3em
The link between the values of the specific intensities at any pair of consecutive
optical depth points $(\tau_{L}, \tau_{L+1})$, namely any single link of the
whole RT chain, is given by the corresponding RT equations in the integral form,
that is

\begin{equation}
I^{+}(\tau _{L},\mu _{J})\ =\ I^{+}(\tau _{L+1},\mu _{J}) \exp
(-{\frac{\Delta \tau _{L}}{\mu _{J}}})\ + \ \int _{\tau _{L}}^{\tau
_{L+1}}S(t) \exp (-{\frac{t-\tau _{L}}{\mu _{J}}})\ dt
\end{equation}  

\parindent=0em
and

\begin{equation}
I^{-}(\tau _{L+1},\mu _{J})\ =\ I^{-}(\tau _{L},\mu _{J})\exp
(-{\frac{\Delta \tau _{L}}{\mu _{J}}})\ +\ \int _{\tau _{L}}^{\tau
_{L+1}}S(t)\exp (-{\frac{\tau _{L+1}-t}{\mu _{J}}})\ dt\ ,
\end{equation}

where $\Delta \tau_{L} \equiv \tau_{L+1} - \tau_{L}$. Equations (6) and (7)
are the straightforward representation of the RT process.

\parindent=3em
At the surface ({\it{i.e.}} for $\tau_{0} = 0$)
the set of values $\lbrace I^{-}(0,\mu_{J}), J = 1,ND\rbrace$ are the 
initial conditions for the inward RT problem, while the set 
$\lbrace I^{+}(0,\mu_{J}), J = 1,ND\rbrace$  is the solution of the 
outward RT problem, {\it{i.e.}} the emergent intensities.
At the bottom the set $\lbrace I^{+}(\tau_{NL},\mu_{J}), J = 1,ND\rbrace$ 
yields the upgoing initial conditions (cf. eq. [5]);
the set 
$\lbrace I^{-}(\tau_{NL},\mu_{J}), J = 1,ND\rbrace$ is the result of the 
inward RT process.

Let us now turn our attention on the cubic spline approximation to the source 
function $S(\tau)$.  That is, we will assume a cubic polynomial approximation
inside each particular interval $(\tau_{L},\tau_{L+1})$,
defined by two consecutive optical depth points,
as the single link of the spline chain. Anyone of these arcs of
cubic is uniquely determined by the values of the
source function and those of its second derivative at 
the end points (knots) of each interval.

To impose the continuity of the source function as well as that of its first
and second derivative at the end points of each interval $(\tau_{L},\tau_{L+1})$
leads to the cubic spline condition

\begin{eqnarray}
\frac{1}{\Delta \tau _{L}}\ S(\tau _{L-1})-(\frac{1}
{\Delta \tau _{L}}+\frac{1}{\Delta \tau
_{L+1}})\ S(\tau _{L})+ \ \frac{1}{\Delta \tau _{L+1}}\ S(\tau _{L+1})\ = 
\nonumber \\ \frac{\Delta \tau_{L}}{6}\ S^{\prime\prime}(\tau _{L-1})+
(\frac{\Delta \tau _{L}}{3}+ \ \frac{\Delta \tau _{L+1}}{3})\ 
S^{\prime\prime}(\tau _{L})+ \frac{\Delta \tau _{L+1}}{6}\ 
S^{\prime\prime}(\tau _{L+1})\ ,
\end{eqnarray}

\parindent=0em
where $\Delta \tau_{L}\ \equiv\ \tau_{L} - \tau_{L-1}$ and
$\Delta \tau_{L+1}\ \equiv\ \tau_{L+1} - \tau_{L}$. Likewise,
as a consequence of the cubic behaviour of $S(\tau)$ between $\tau_{L}$
and $\tau_{L+1}$, it will hold that
\begin{equation}
S^{\prime}(\tau _{L})\ =\ \frac{1}{\Delta \tau _{L}}\ [S(\tau _{L+1})-S(\tau
_{L})]-\ \frac{\Delta \tau _{L}}{3}S^{\prime\prime}(\tau _{L})-
\frac{\Delta \tau _{L}}{6}S^{\prime\prime}(\tau _{L+1})\ ,
\end{equation}

\begin{equation}
S^{\prime}(\tau _{L+1})\ =\ \frac{1}{\Delta \tau _{L}}\ [S(\tau _{L+1})-S(\tau
_{L})]+\ \frac{\Delta \tau _{L}}{6}S^{\prime\prime}(\tau _{L})+
\frac{\Delta \tau _{L}}{3}S^{\prime\prime}(\tau _{L+1})
\end{equation}
and

\begin{equation}
S^{\prime\prime\prime}(\tau _{L})\ =\ \frac{1}{\Delta \tau _{L}}\ 
[S^{\prime\prime}(\tau _{L+1})-S^{\prime\prime}(\tau _{L})]\ .
\end{equation}
\bigskip

By means of eq. (8) we are in a position to join the neighbouring links of the
spline chain while ensuring the required continuity at the knots.

\parindent=3em
Like for the RT chain, also for the cubic spline chain we need two bundary 
conditions. Customarily these are 
$S^{\prime\prime}(\tau_{0}) = S^{\prime\prime}(\tau_{1})$ and
$S^{\prime\prime}(\tau_{NL}) = S^{\prime\prime}(\tau_{NL-1})$. In the present 
study we assume that $S^{\prime\prime}(\tau_{0}) = S^{\prime\prime}(\tau_{1})$
at the surface, namely that the first arc of the spline chain is a parabola. 
On the contrary, the boundary condition for the spline chain at the bottom
must be consistent with the diffusion approximation for the incident upgoing
intensities $I^{+}(\tau_{NL},\mu_{J})$, given by eq. (5), which is a 
consequence of having assumed also a cubic 
polynomial behaviour for $S(\tau)$ at depths greater than $\tau_{NL}$.
This condition is in agreement with the cubic polynomial behaviour of $S(\tau)$
inside the last layer $(\tau_{NL-1},\tau_{NL})$. Hence we cannot introduce now
a different approach to $S(\tau)$.
However we can derive the formal value of the first derivative
$S^{\prime}(\tau)$ from eq. (2), that is

\begin{equation}
S^{\prime}(\tau)\ =\ [\varepsilon(\tau) B(\tau)]\ ^{\prime}\ +\ 
[1 - \varepsilon(\tau)]\ ^{\prime}\ J(\tau)\ +\ [1 - \varepsilon(\tau)]\
J^{\prime}(\tau)\ ,
\end{equation}

\parindent=0em
where
\begin{equation}
J^{\prime}(\tau_{L})\ =\ \sum_{J=1}^{ND}\ {\bf{wd}}_{J}\
\frac{I^{+}(\tau_{L},\mu_{J})\ -\ I^{-}(\tau_{L},\mu_{J})}{\mu_{J}}\ .
\end{equation}

Equation (12), evaluated at the deepest optical depth point $\tau_{NL}$ will 
then yield the required lower boundary condition, as will be shown later.

\parindent=3em
Let us get back now to eq.s (6) and (7). For any interval $(\tau_{L},
\tau_{L+1})$ the arc of cubic approximating to $S(\tau)$ is given by

\begin{equation}
S(\tau )\ =\ S(\tau _{L})+S^{\prime}(\tau _{L})(\tau -\tau _{L})+
\frac{1}{2}S^{\prime\prime}(\tau _{L}) (\tau -\tau _{L})^{2}+
\frac{1}{6}S^{\prime\prime\prime}(\tau _{L})(\tau -\tau _{L})^{3} ,
\end{equation}

\parindent=0em
By replacing eq. (14) in eq.s (6) and (7), and taking into account eq.s
(9) through (11),  we get eventually
 
\begin{eqnarray}\nonumber
I^{+}(\tau _{L},\mu _{J})\ =\ I^{+}(\tau _{L+1},\mu
_{J})\exp -(\frac{\Delta \tau _{L}}{\mu _{J}})+ 
\end{eqnarray}
\begin{equation}
{\bf{ws}}_{1}(J)\ S(\tau
_{L})+  {\bf{ws}}_{2}(J)\ S(\tau _{L+1})+ \nonumber \\ {\bf{wd}}_{1}(J)\ S''(\tau
_{L})+ {\bf{wd}}_{2}(J)\ S''(\tau _{L+1})
\end{equation}

and

\begin{eqnarray}\nonumber
I^{-}(\tau _{L+1},\mu _{J})\ =\ I^{-}(\tau _{L},\mu
_{J})\exp -(\frac{\Delta \tau _{L}}{\mu _{J}})+ 
\end{eqnarray}
\begin{equation}
{\bf{ws}}_{2}(J)\ S(\tau
_{L})+{\bf{ws}}_{1}(J)\ S(\tau _{L+1})+ \nonumber \\{\bf{wd}}_{2}(J)\ S''(\tau
_{L})+{\bf{wd}}_{1}(J)\ S''(\tau _{L+1}) .
\end{equation}

The quadrature weights ${\bf{ws}}_{1}(J), {\bf{ws}}_{2}(J), {\bf{wd}}_{1}(J)$ 
and ${\bf{wd}}_{2}(J)$ are computed straightforwardly by taking into account
eq.s (9) through (11) to yield

\begin{equation}
{\bf{ws}}_{1}(J)\ =\ [1-\frac{1}{\delta }]+\frac{1}{\delta }\ e^{-\delta}\ ,
\end{equation}

\begin{equation}
{\bf{ws}}_{2}(J)\ =\ \frac{1}{\delta }-[1+\frac{1}{\delta }]\ e^{-\delta}\ ,
\end{equation}

\begin{equation}
{\bf{wd}}_{1}(J)\ =\ \mu ^{2}\ [(1-\frac{\delta }{3}-\frac{1}{\delta})-
(\frac{\delta }{6}-\frac{1}{\delta })\ e^{-\delta }]\ ,
\end{equation}

\begin{equation}
{\bf{wd}}_{2}(J)\ =\ \mu ^{2}\ [(-{\frac{\delta }{6}}+\frac{1}{\delta})-
(1+\frac{\delta }{3}+\frac{1}{\delta })\ e^{-\delta }]\ , 
\end{equation}

where $\delta\ =\ \Delta \tau_{L}\ /\ \mu_{J}$.

\parindent=3em
Sometimes, when $\delta << 1$, for sake of numerical percision it may be 
necessary to recast the foregoing weights into the form

\begin{equation}
{\bf{ws}}_{1}(J)\ =\ \frac{1}{2} \delta\ -\ \frac{1}{6} \delta^{2}\ +\ 
\frac{1}{24} \delta^{3}\ -\ \frac{1}{120} \delta^{4}\ +\ \frac{1}{720} 
\delta^{5}\ -\ \frac{1}{5040} \delta^{6}º +\ \frac{1}{40320} \delta^{7}\ ,
\end{equation}

\begin{equation}
{\bf{ws}}_{2}(J)\ =\ \frac{1}{2} \delta\ -\ \frac{1}{3} \delta^{2}\ +\ 
\frac{1}{8} \delta^{3}\ -\ \frac{1}{30} \delta^{4}\ +\ \frac{1}{144} 
\delta^{5}\ -\ \frac{1}{840} \delta^{6}º +\ \frac{1}{5760} \delta^{7}\ ,
\end{equation}

\begin{equation}
{\bf{wd}}_{1}(J)\ =\ -\ \mu^{2}\ 
(\frac{1}{24} \delta^{3}\ -\ \frac{7}{360} \delta^{4} 
\ +\ \frac{1}{180} \delta^{5}\ -\ \frac{1}{840} \delta^{6}\ +\ \frac{5}{24192} 
\delta^{7})\ ,
\end{equation}

\begin{equation}
{\bf{wd}}_{2}(J)\ =\ -\ \mu^{2}\ 
(\frac{1}{24} \delta^{3}\ -\ \frac{1}{45} \delta^{4} 
\ +\ \frac{1}{144} \delta^{5}\ -\ \frac{1}{630} \delta^{6}\ +\ \frac{1}{3456} 
\delta^{7})\ .
\end{equation}
\bigskip

To conclude, eq.s (15) and (16) together with eq.s (17) through (20) allow us to
write explicitly for each direction $\mu_{J}$ the relations between 
$I^{+}(\tau_{L},\mu_{J})$ and $I^{+}(\tau_{L+1},\mu_{J})$
on the one hand, between $I^{-}(\tau_{L+1},\mu_{J})$
and $I^{-}(\tau_{L},\mu_{J})$ on the other. These relations are linear functions
of the unknown values of  
$S(\tau_{L})$, $S(\tau_{L+1})$, $S^{\prime\prime}(\tau_{L})$ and 
$S^{\prime\prime}(\tau_{L+1})$, which will play a protagonist role in the 
numerical algorithm. The cubic spline
condition, given by eq. (8), impose a further relation between $S(\tau_{L})$
and $S^{\prime\prime}(\tau_{L})$ at each knot $\tau_{L}$. 

We recall that
for any frequency the specific source function is approximated by
an arc of cubic inside each interval $(\tau_{L},\tau_{L+1})$. Therefore
in the layers 
deeper than the corresponding spectral formation region,
where $\exp \lbrace - \Delta\tau\rbrace$ is pratically null, the form of the
weights ${\bf{ws}}_{1}(J)$, ${\bf{ws}}_{2}(J)$, ${\bf{wd}}_{1}(J)$ and
${\bf{wd}}_{2}(J)$ given by eq.s (17) through (20) warrants that the intensities
$I^{+}(\tau,\mu_{J})$ recover there the form of eq. (5), originally assigned at
the bottom of the atmosphere ({\it{i.e.}} at $\tau_{NL}$). That is to say, the
boundary condition, initially assigned at the bottom, is transported up to the
end of the spectral formation region, keeping its form in a natural way.
On the other hand, in the outer layers beyond the region of 
formation,where $\exp \lbrace - \Delta\tau\rbrace$ approaches unity,
eq.s (6) and (7) warrant
that $I^{+}(\tau,\mu_{J})$ and $I^{-}(\tau,\mu_{J})$ keep constant. Thus, albeit
the total number $NL$ of layers exceed that required by the proper physical
treatment of the formation region for each single frequency, such an excess
does not affect the numerical computation of the protagonist variables.
That is to say, frequency by frequency the effective transport of the specific
intensities is performed in a natural way inside its own region of formation, 
provided that care has be taken to select the geometrical width of the stellar
atmosphere system so that, as already stressed in the Introduction, the former 
include the region of formation for all the frequencies.

We have then at hand 
all the mathematical tools that will allow us to solve the global RT problem in
the same way as in the original IIM scheme (see Paper I).

However only to warrant the continuity of the two first derivatives of the
source function is not enough to avoid the occurence of instabilities. As in the
cubic spline fundamental equation (eq. [8])
the protagonist variables are the function
itself and its second derivative (both tied through their values at any set of
three consecutive points), also in the RT elimination scheme the source function
and its second derivative must be the protagonist variables.

In a previous attempt we formulated the equations (15) and (16),
which describe the propagation of the upgoing and downgoing intensities, in 
terms of $S(\tau_{L})$, $S(\tau_{L+1})$, $S^{\prime}(\tau_{L})$ and 
$S^{\prime}(\tau_{L+1})$ after the elimination of 
$S^{\prime\prime}(\tau_{L})$ and $S^{\prime\prime}(\tau_{L+1})$ given as 
functions of $S(\tau_{L})$, $S(\tau_{L+1})$, $S^{\prime}(\tau_{L})$ and 
$S^{\prime}(\tau_{L+1})$ thanks to the cubic behaviour of $S(\tau)$.
In the  actual
version we describe the propagation of the aforesaid intensities by means
of $S(\tau_{L})$, $S(\tau_{L+1})$, $S^{\prime\prime}(\tau_{L})$ and 
$S^{\prime\prime}(\tau_{L+1})$, again by means of the cubic behaviour of 
$S(\tau)$.
From the mathematical point of view both representations should yield the same 
results, but from the numerical standpoint it looks much better to work directly
with the second derivatives   
$S^{\prime\prime}(\tau_{L})$ and $S^{\prime\prime}(\tau_{L+1})$, because the 
fundamental equation (8), that links the sequence of succesive layers in the
cubic spline scheme, requires the variables 
$S(\tau)$ and $S^{\prime\prime}(\tau)$. 

In the present formulation of the propagation equations (15) and (16) the
integration weights 
${\bf{ws}}_{1}(J)$ and ${\bf{ws}}_{2}(J)$, given by eq.s (17) and (18), account
strictly for the linear piece-wise approximation to any monochromatic source
function $S_{\nu}(\tau)$. The remaining weights,
${\bf{wd}}_{1}(J)$ and ${\bf{wd}}_{2}(J)$, account for the deviation from the 
linear behaviour, either parabolic or cubic. Whenever  
${\bf{wd}}_{1}(J)$ and ${\bf{wd}}_{2}(J)$ take on small values, the linear
approximation is more than enough. This is the case in the outermost layers,
where it holds that $\delta < 1$ and $\delta^{3} << 1$; the linear approximation
is automatically recovered, as only the weights
${\bf{ws}}_{1}(J)$ and ${\bf{ws}}_{2}(J)$ account for the variation of the
source function in optically thin layers. That is to say, in the practice only
$S(\tau_{L})$ and $S(\tau_{L+1})$ take part in the elimination scheme. In other
words, the effects of a non-linear behaviour play the role of a perturbation of
the linear behaviour.

In the original formulation of the IIM (Paper I), we employed the variables
$S(\tau_{L})$, $S(\tau_{L+1})$, $S^{\prime}(\tau_{L})$ and 
$S^{\prime}(\tau_{L+1})$, together with the corresponding integration weights,
in order to describe the propagation of the upgoing and downgoing intensities
between any pair of optical depth points $\tau_{L}$ and $\tau_{L+1}$. Whatever
their behaviour (linear, quadratic or cubic), all the four variables and the
relevant integration weights took an active part in the elimination scheme, both
from the theoretical and the numerical standpoint. This can have been at the
origin of the instabilities that showed up, above all in the regions of small 
optical depth.
The actual version of the IIM, due to the above mentioned reasons, results 
certainly more reliable.

\section{The Forward-Elimination/Back-Substitution scheme}

As already said, we will work with a set of fundamental variables whose values
are unknown: the upgoing and downgoing specific intensities
$I^{\pm}(\tau_{L},\mu_{J})$, the corresponding source
functions $S(\tau_{L})$ and their second derivatives 
$S^{\prime\prime}(\tau_{L})$. The major aim of this section is to
derive linear
relations among the values of the foregoing fundamental variables at the two
consecutive optical depth points $\tau_{L}$ and $\tau_{L+1}$ that delimitate 
each of the layer $(\tau_{L},\tau_{L+1})$ succesively
under study. The coefficients of these relations 
are easily computed, and will be denoted in the following by bold face symbols.

\subsection{The algorithmic representation of the upper boundary conditions}

We start necessarily with only one half of the data of the problem, namely
the set of the downgoing intensities incident onto the upper boundary layer
at $\tau_{0} = 0$,  {\it{i.e.}}
$\lbrace I^{-}(\tau_{0},\mu_{J}), J = 1,ND\rbrace$, 
that we will write in its most general form as
\bigskip

\begin{eqnarray}\nonumber
I^{-}(\tau_{0},\mu_{J})\ =\ {\bf{cm0}}(J)\ +\ {\bf{cms1}}(J)\ S(\tau_{0})\ +\ 
{\bf{cms2}}(J)\ S(\tau_{1})\ + 
\end{eqnarray}
\begin{equation}
{\bf{cmds1}}(J)\ S^{\prime\prime}(\tau_{0})\ +
{\bf{cmds2}}(J)\ S^{\prime\prime}(\tau_{1})\ +\ 
\sum_{J^{\prime} = 1}^{ND}\ {\bf{R}}(J,J^{\prime})\ 
I^{+}(\tau_{0},\mu_{J^{\prime}})\ .
\end{equation}

\parindent=0em
The coefficient ${\bf{cm0}}(J)$ accounts for the numerical value of the incident
intensity $I^{-}(\tau_{0},\mu_{J})$, which is usually null. The reflexion matrix
${\bf{R}}(J,J^{\prime})$ takes into account the possible effects of 
backscattering outside the stellar surface. Under usual conditions it holds that
also ${\bf{R}}(J,J^{\prime}) = 0$. 
On physical grounds it is hard to justify the dependence of 
$I^{-}(\tau_{0},\mu_{J})$ on the values of the source function and its second
derivative at points $\tau_{0}$ and $\tau_{1}$ through the coefficients
${\bf{cms1}}(J), {\bf{cms2}}(J), {\bf{cmds1}}(J)$ and ${\bf{cmds2}}(J)$. It is
rather an algorithmical requirement, as these coefficients allow us to link
linearly the values of the protagonist variables between two consecutive optical
depth points. Consistently with the upper boundary conditions, the latter 
coefficients have to be set equal to zero.

\parindent=3em
At the end of the treatment of radiative transfer in the first layer (as well as
in the succesive ones) some of these coefficients will take on values 
different from 
zero. These new values can overrun the previous memory storage, because the 
current relation for the downgoing intensities at $\tau_{0} = 0$
will not be necessary any longer. 

Inside the forward-elimination scheme for the RT process we must propagate not
only the upgoing and downgoing specific intensities (which brings about the
propagation of the source function as defined by eq.s (2) and (3)), but also
the second derivative $S^{\prime\prime}(\tau)$ of the source function in the
cubic spline scheme.

As already said, we assume that in the first layer $(\tau_{0},\tau_{1})$ 
the source function $S(\tau)$
can be approximated by an arc of parabola, which implies that
$S^{\prime\prime}(\tau_{0}) = S^{\prime\prime}(\tau_{1})$. This condition will
be included in the coefficients of the relation
\begin{eqnarray}\nonumber
S^{\prime\prime}(\tau_{0})\ =\ {\bf{cds0}}\ +\ {\bf{cds1}}\ S(\tau_{0})\ +\
{\bf{cds2}}\ S(\tau_{1})\ +
\end{eqnarray}
\begin{equation}
{\bf{cdds1}}\ S^{\prime\prime}(\tau_{0})\ +\ 
{\bf{cdds2}}\ S^{\prime\prime}(\tau_{1})\ +\ 
\sum_{J = 1}^{ND}\ {\bf{cdi}}(J)\ \ I^{+}(\tau_{0},\mu_{J})\ ,
\end{equation}

\parindent=0em
where the values of 
$S(\tau_{0}), S(\tau_{1}), S^{\prime\prime}(\tau_{0}), 
S^{\prime\prime}(\tau_{1})$ and  the set
$\lbrace I^{+}(\tau_{0},\mu_{J}), J = 1,ND\rbrace$ are
unknown. In order to fulfill the above boundary condition, all the 
coefficients in eq. (26) must be equal to zero, excepted
${\bf{cdss2}}$ that must be
set equal to one. To express here $S^{\prime\prime}(\tau_{0})$ as a function of
$S(\tau_{0})$, $S(\tau_{1})$, $S^{\prime\prime}(\tau_{0})$ itself, 
$S^{\prime\prime}(\tau_{1})$ and the set
$\lbrace I^{+}(\tau_{0},\mu_{J}), J = 1,ND\rbrace$
is just for algorithmical ease. When convenient,
we will solve for $S(\tau_{0})$ - and for $S^{\prime\prime}(\tau_{0})$ -
in terms of $S(\tau_{1})$, $S^{\prime\prime}(\tau_{1})$ and 
$\lbrace I^{+}(\tau_{1},\mu_{J})\rbrace$.

\subsection{The layer by layer elimination}

We are going to show
here how the treatment of the first layer $(\tau_{0},\tau_{1})$,
labelled by $L = 1$, will yield the coefficients of the relation
\begin{eqnarray}\nonumber
S(\tau_{0})\ =\ {\bf{cbs0}}(1)\ +
\end{eqnarray}
\begin{equation}
{\bf{cbss}}(1)\ S(\tau_{1})\ +\  
{\bf{cbsd}}(1)\ S^{\prime\prime}(\tau_{1})\ +\  
\sum_{J = 1}^{ND}\ {\bf{cbsi}}(1,J)\ \ I^{+}(\tau_{1},\mu_{J})
\end{equation}

\parindent=0em
and those of the relation
\begin{eqnarray}\nonumber
S^{\prime\prime}(\tau_{0})\ =\ {\bf{cbd0}}(1)\ +
\end{eqnarray}
\begin{equation}
{\bf{cbds}}(1)\ S(\tau_{1})\ +\  
{\bf{cbdd}}(1)\ S^{\prime\prime}(\tau_{1})\ +\  
\sum_{J = 1}^{ND}\ {\bf{cbdi}}(1,J)\ \ I^{+}(\tau_{1},\mu_{J})\ .
\end{equation}

These coefficients will be stored in order to compute $S(\tau_{0})$ and 
$S^{\prime\prime}(\tau_{0})$ in the succesive back-substitution process, once
the values of
$S(\tau_{1})$, $S^{\prime\prime}(\tau_{1})$  as well as the set
$\lbrace I^{+}(\tau_{1},\mu_{J}), J = 1,ND\rbrace$ have been determined. The 
above relations link any pair of succesive layers. As already said,
the determination of these relations constitutes the aim of this section.

\parindent=3em
In parallel we are going to show also how
to recover the initial conditions for $I^{-}(\tau_{1},\mu_{J})$ and 
$S^{\prime\prime}(\tau_{1})$, {\it{i.e.}} the values of the coefficients of the
relations equivalent to eq.s (25) and (26), now for $\tau_{1}$. 

Let us detail our foregoing purpose.
At the beginning of the study of each succesive layer - here the first one -
we must consider the implicit computation of the corresponding source function
at the upper limiting optical depth, here $\tau_{0}$. The form 
of the incident downgoing intensities  at $\tau_{0}$, given by eq. (25),
together with the implicit values of the set 
$\lbrace I^{+}(\tau_{0},\mu_{J}), J = 1,ND\rbrace$ allow us to compute from eq. (4) the
coefficients of a linear relation among $J(\tau_{0})$ and
$S(\tau_{0}), S(\tau_{1}), S^{\prime\prime}(\tau_{0}), 
S^{\prime\prime}(\tau_{1})$ and the set
$\lbrace I^{+}(\tau_{0},\mu_{J}), J = 1,ND\rbrace$. Then eq. (2),  where
$\varepsilon(\tau_{0})$ and $B(\tau_{0})$ are given, will yield the coefficients
 of the linear relation 
\begin{eqnarray}\nonumber
S(\tau_{0})\ =\ {\bf{cs0}}\ +\ {\bf{css1}}\ S(\tau_{0})\ +\ 
{\bf{css2}}\ S(\tau_{1})\ +
\end{eqnarray}
\begin{equation}
{\bf{csds1}}\ S^{\prime\prime}(\tau_{0})\ +\ 
{\bf{csds2}}\ S^{\prime\prime}(\tau_{1})\ +\ 
\sum_{J = 1}^{ND}\ {\bf{csi}}(J)\ I^{+}(\tau_{0},\mu_{J})\ .
\end{equation}

\parindent=0em
where we have not solved for $S(\tau_{0})$ again for the sake of 
algorithmical ease.

\parindent=3em
We compute now for each direction $\mu_{J}$ the quadrature weights
${\bf{ws1}}(J)$, ${\bf{ws2}}(J)$, ${\bf{wd1}}(J)$ and ${\bf{wd2}}(J)$ 
according to
eq.s (17) through (20) - or alternatively eq.s (21) through (24) - for
$\Delta\tau_{1} = \tau_{1} - \tau_{0}$. These weights allow us an implicit 
quadrature of the source function in the description of the propagation of the 
upgoing intensities from $I^{+}(\tau_{1},\mu_{J})$ to $I^{+}(\tau_{0},\mu_{J})$,
and later of the downgoing intensities from 
$I^{-}(\tau_{0},\mu_{J})$ to $I^{-}(\tau_{1},\mu_{J})$.

At this point we can introduce the implicit form for $I^{+}(\tau_{0},\mu_{J})$
in terms of
$S(\tau_{0})$, $S(\tau_{1})$, $S^{\prime\prime}(\tau_{0})$, 
$S^{\prime\prime}(\tau_{1})$  and the set
$\lbrace I^{+}(\tau_{1},\mu_{J}), J = 1,ND\rbrace$,
given by eq. (15), in eq. (25) for $I^{-}(\tau_{0},\mu_{J})$,
which is the initial condition for the study 
of the layer $(\tau_{0},\tau_{1})$. By re-arrangement of
the coefficients we can write
\begin{eqnarray}\nonumber
I^{-}(\tau_{0},\mu_{J})\ =\ {\bf{cm0}}(J)\ +\ {\bf{cms1}}(J)\ S(\tau_{0})\ +\ 
{\bf{cms2}}(J)\ S(\tau_{1})\ + 
\end{eqnarray}
\begin{equation}
{\bf{cmds1}}(J)\ S^{\prime\prime}(\tau_{0})\ +
{\bf{cmds2}}(J)\ S^{\prime\prime}(\tau_{1})\ +\ 
\sum_{J^{\prime} = 1}^{ND}
\ {\bf{R}}(J,J^{\prime})\ I^{+}(\tau_{1},\mu_{J^{\prime}})\ .
\end{equation}

\parindent=0em
for any direction $\mu_{J}$. These new values of the coefficients can 
overrun the
memory places of the pervios ones, corresponding to the intial condition given
by eq. (25).

\parindent=3em
We repeat the same exercice, namely to employ eq. (15) inside both the 
functional form for $S(\tau_{0})$, given by eq. (29), and that for
$S^{\prime\prime}(\tau_{0})$, given by eq. (26), in order to recover the
previous form for both of them, but now as a function of the upgoing 
intensities at 
$\tau_{1}$ insted of $\tau_{0}$, hence with different coeffficients. That is
\begin{eqnarray}\nonumber
S(\tau_{0})\ =\ {\bf{cs0}}\ +\ {\bf{css1}}\ S(\tau_{0})\ +\ 
{\bf{css2}}\ S(\tau_{1})\ +
\end{eqnarray}
\begin{equation}
{\bf{csds1}}\ S^{\prime\prime}(\tau_{0})\ +\ 
{\bf{csds2}}\ S^{\prime\prime}(\tau_{1})\ +\ 
\sum_{J = 1}^{ND}\ {\bf{csi}}(J)\ I^{+}(\tau_{1},\mu_{J})\ .
\end{equation}

\parindent=0em
and
\begin{eqnarray}\nonumber
S^{\prime\prime}(\tau_{0})\ =\ {\bf{cds0}}\ +\ {\bf{cds1}}\ S(\tau_{0})\ +\
{\bf{cds2}}\ S(\tau_{1})\ +
\end{eqnarray}
\begin{equation}
{\bf{cdds1}}\ S^{\prime\prime}(\tau_{0})\ +\ 
{\bf{cdds2}}\ S^{\prime\prime}(\tau_{1})\ +\ 
\sum_{J = 1}^{ND}\ {\bf{cdi}}(J)\ \ I^{+}(\tau_{1},\mu_{J})\ .
\end{equation}

Now, just by solving for 
$S(\tau_{0})$ and $S^{\prime\prime}(\tau_{0})$ we obtain the coefficients of the
relations (27) and (28), earlier announced at the beginning of Section 3.2. 
These coefficients must be stored for further use.  In such a way we have 
achieved part of out goal.

\parindent=3em
At this point let us describe the propagation of the downgoing intensities
from $I^{-}(\tau_{0},\mu_{J})$, given by eq. (30), to $I^{-}(\tau_{1},\mu_{J})$
according to eq. (16). By re-arrangement of the coefficients we get the 
new values corresponding to the relation
\begin{eqnarray}\nonumber
I^{-}(\tau_{1},\mu_{J})\ =\ {\bf{cm0}}(J)\ +\ {\bf{cms1}}(J)\ S(\tau_{0})\ +\ 
{\bf{cms2}}(J)\ S(\tau_{1})\ + 
\end{eqnarray}
\begin{equation}
{\bf{cmds1}}(J)\ S^{\prime\prime}(\tau_{0})\ +
{\bf{cmds2}}(J)\ S^{\prime\prime}(\tau_{1})\ +\ 
\sum_{J^{\prime} = 1}^{ND}
\ {\bf{R}}(J,J^{\prime})\ I^{+}(\tau_{1},\mu_{J^{\prime}})
\end{equation}

\parindent=0em
for all the directions $\mu_{J}$.
Again these new values of the coefficients can overrun the previous ones,
corresponding to eq. (30). 

\parindent=3em
If we introduce the foregoing
eq.s (27) and (28), whose coefficients we have just computed,
in the functional form of $I^{-}(\tau_{1},\mu_{J})$ given by eq. (33), by
re-arrangement of the previous coefficients we derive the new ones for 
the relation
\begin{eqnarray}\nonumber
I^{-}(\tau_{1},\mu_{J})\ =
\end{eqnarray}
\begin{equation}
{\bf{cm0}}(J)\ +\ {\bf{cms1}}(J)\ S(\tau_{1})\ +\ 
{\bf{cmds1}}(J)\ S^{\prime\prime}(\tau_{1})\ +
\sum_{J^{\prime} = 1}^{ND}\ 
{\bf{R}}(J,J^{\prime})\ I^{+}(\tau_{1},\mu_{J^{\prime}})\ ,
\end{equation}

\parindent=0em
which we will cast into the form required by eq. (25) by setting equal to zero 
the coefficients ${\bf{cms2}}(J)$ and ${\bf{cmds2}}(J)$. We have thus determined
the coefficients of the linear relation required as the initial condition at
$\tau_{1}$, that will be necessary to study the propagation of the downgoing
intensities in the succesive layer $(\tau_{1},\tau_{2})$.

\parindent=3em
We have still to determine the initial condition for the propagation of
$S^{\prime\prime}(\tau)$, that is to say a linear relation like eq. (26),
now for $\tau_{1}$. It is matter of recovering the functional form of
$S^{\prime\prime}(\tau_{1})$ in ordert to start the study of the spline chain
in the layer $(\tau_{1},\tau_{2})$. We have at hand the fundamental relation
for the cubic spline, namely eq. (8) that links linearly
$S^{\prime\prime}(\tau_{0})$, $S^{\prime\prime}(\tau_{1})$ and 
$S^{\prime\prime}(\tau_{2})$ with $S(\tau_{0})$, $S(\tau_{1})$ and 
$S(\tau_{2})$.

By introducing in eq. (8) the formal expressions for $S(\tau_{0})$ and
$S^{\prime\prime}(\tau_{0})$, given by eq.s (27) and (28),
we get easily the coefficients of the equation
\begin{eqnarray}\nonumber
S^{\prime\prime}(\tau_{1})\ =\ {\bf{cds0}}\ +\ {\bf{cds1}}\ S(\tau_{1})\ +\
{\bf{cds2}}\ S(\tau_{2})\ +
\end{eqnarray}
\begin{equation}
{\bf{cdds1}}\ S^{\prime\prime}(\tau_{1})\ +\ 
{\bf{cdds2}}\ S^{\prime\prime}(\tau_{2})\ +\ 
\sum_{J = 1}^{ND}\ {\bf{cdi}}(J)\ \ I^{+}(\tau_{1},\mu_{J})\ ,
\end{equation}

\parindent=0em
akin to eq. (26), the bootstrap at $\tau_{0}$, that was the initial condition
to studying the layer $(\tau_{0},\tau_{1})$. Equation (35), together with
eq. (33) that is the initial condition for the treatment of radiative transfer,
will allow us to repeat the foregoing procedure for the layer 
$(\tau_{1},\tau_{2})$. This scheme is then iterated layer by layer till the 
bottom of the atmosphere.

\subsection{The solution at the bottom and the Back-Substitution}

\parindent=3em
At the end of the forward-elimination scheme we have at hand the full set
of coefficients of eq.s (27) and (28) for each optical depth of the set
$\lbrace \tau_{0}, \tau_{1}, ... ,\tau_{NL-1}\rbrace$. The explicit values of
$S(\tau_{L})$ and $S^{\prime\prime}(\tau_{L})$ as well as those of the set of 
the outgoing intensities $\lbrace I^{+}(\tau_{L},\mu_{J}), J=1,ND\rbrace$ 
have now to be computed in the back-substitution scheme.

For the sake of a more clear exposition of the mathematical solution at the 
bottom we will rewrite eq.s (27) and (28) for $\tau_{NL}$, that is

\begin{eqnarray}\nonumber
S(\tau_{NL-1})\ =\ {\bf{cbs0}}(NL-1)\ +
\end{eqnarray}
\begin{equation}
{\bf{cbss}}(NL-1)\ S(\tau_{NL})\ +\  
{\bf{cbsd}}(NL-1)\ S^{\prime\prime}(\tau_{NL})\ +\  
\sum_{J = 1}^{ND}\ {\bf{cbsi}}(NL-1,J)\ \ I^{+}(\tau_{NL},\mu_{J})
\end{equation}

\parindent=0em
and
\begin{eqnarray}\nonumber
S^{\prime\prime}(\tau_{NL-1})\ =\ {\bf{cbd0}}(NL-1)\ +
\end{eqnarray}
\begin{equation}
{\bf{cbds}}(NL-1)\ S(\tau_{NL})\ +\  
{\bf{cbdd}}(NL-1)\ S^{\prime\prime}(\tau_{NL})\ +\  
\sum_{J = 1}^{ND}\ {\bf{cbdi}}(NL-1,J)\ \ I^{+}(\tau_{NL},\mu_{J})\ .
\end{equation}

Also, at the end of the forward-elimination scheme, the current values of the
coefficients of the equation for $I^{-}(\tau_{NL},\mu_{J})$, that is
\begin{eqnarray}\nonumber
I^{-}(\tau_{NL},\mu_{J})\ =\ {\bf{cm0}}(J)\ +\ {\bf{cms1}}(J)\ S(\tau_{NL})\ +\ 
{\bf{cms2}}(J)\ S(\tau_{NL+1})\ + 
\end{eqnarray}
\begin{equation}
{\bf{cmds1}}(J)\ S^{\prime\prime}(\tau_{NL})\ +
{\bf{cmds2}}(J)\ S^{\prime\prime}(\tau_{NL+1})\ +\ 
\sum_{J^{\prime}}\ {\bf{R}}(J,J^{\prime})\ I^{+}(\tau_{NL},\mu_{J^{\prime}})\ ,
\end{equation}

are still stored in the scratch memory. For the sake of a homogeneous algorithm 
we had kept the dependence on $S(\tau_{NL+1})$ and 
$S^{\prime\prime}(\tau_{NL+1})$ through the coefficients ${\bf{cms2}}(J)$ and
${\bf{cmds2}}(J)$. But these coefficients are null so that $S(\tau_{NL+1})$ and
$S^{\prime\prime}(\tau_{NL+1})$ do not play any active role. The same 
algorithmical requirement compelled us to introduce the dummy supplementary
optical depth $\tau_{NL+1}$.

\parindent=3em
At this point we can apply the lower boundary condition for the radiative
transfer, {\it{i.e.}} the formal expression for $I^{+}(\tau_{NL},\mu_{J})$ 
given by
eq. (5). If we replace this expression in the previous eq.s (36), (37) and (38),
by re-arrangement of terms we obtain the explicit values of the
coefficients of the two linear relations for $S(\tau_{Nl-1})$ and
$S^{\prime\prime}(\tau_{NL-1})$ as a function of
$S(\tau_{NL})$, $S^{\prime}(\tau_{NL})$, $S^{\prime\prime}(\tau_{NL})$ and
$S^{\prime\prime\prime}(\tau_{NL})$, that we will write as
\begin{equation}
S(\tau_{NL-1})\ =\ {\cal{LR}}\lbrace S(\tau_{NL}),S^{\prime}(\tau_{NL}),
S^{\prime\prime}(\tau_{NL}),S^{\prime\prime\prime}(\tau_{NL})\rbrace
\end{equation} 

\parindent=0em
and
\begin{equation}
S^{\prime\prime}(\tau_{NL-1})\ =
\ {\cal{LR}}\lbrace S(\tau_{NL}),S^{\prime}(\tau_{NL}),
S^{\prime\prime}(\tau_{NL}),S^{\prime\prime\prime}(\tau_{NL})\rbrace\ .
\end{equation} 

Likewise, if we take into account the aforesaid expression for 
$I^{-}(\tau_{NL},\mu_{J})$, whose coefficients ${\bf{cms2}}(J)$ and
${\bf{cmds2}}(J)$ are null, we get also the coefficients of the linear
relations
\begin{equation}
I^{-}(\tau_{NL},\mu_{J})\ =
\ {\cal{LR}}\lbrace S(\tau_{NL}),S^{\prime}(\tau_{NL}),
S^{\prime\prime}(\tau_{NL}),S^{\prime\prime\prime}(\tau_{NL})\rbrace
\end{equation} 

for each direction $\mu_{J}$.

\parindent=3em
In the forward-elimination, at the beginning of the study of each layer 
$(\tau_{L},\tau_{L+1})$, we have formally computed the mean intensity 
$J(\tau_{L})$ and the corresponding source function $S(\tau_{L})$ at the upper
optical depth $\tau_{L}$. That is to say, we have not yet used the relation 
given by eq. (2) at the last optical depth $\tau_{NL}$. Let us do it here.

\parindent=3em
Equation (5) for $I^{+}(\tau_{NL},\mu_{J})$ and (41) for 
$I^{-}(\tau_{NL},\mu_{J})$ allow us to compute via eq. (4)
the coefficients of a linear relation like
\begin{equation}
J(\tau_{NL})\ =\ {\cal{LR}}\lbrace S(\tau_{NL}),S^{\prime}(\tau_{NL}),
S^{\prime\prime}(\tau_{NL}),S^{\prime\prime\prime}(\tau_{NL})\rbrace
\end{equation} 

\parindent=0em
Now thanks to eq. (13) we can compute also the coefficients of the linear 
relation
\begin{equation}
J^{\prime}(\tau_{NL})\ =\ {\cal{LR}}\lbrace S(\tau_{NL}),S^{\prime}(\tau_{NL}),
S^{\prime\prime}(\tau_{NL}),S^{\prime\prime\prime}(\tau_{NL})\rbrace\ .
\end{equation}

By means of eq.s (2) and (12)
we can eventually derive the explicit coefficients of the linear relations
\begin{equation}
S(\tau_{NL})\ =\ {\cal{LR}}\lbrace S(\tau_{NL}),S^{\prime}(\tau_{NL}),
S^{\prime\prime}(\tau_{NL}),S^{\prime\prime\prime}(\tau_{NL})\rbrace
\end{equation} 

and
\begin{equation}
S^{\prime}(\tau_{NL})\ =\ {\cal{LR}}\lbrace S(\tau_{NL}),S^{\prime}(\tau_{NL}),
S^{\prime\prime}(\tau_{NL}),S^{\prime\prime\prime}(\tau_{NL})\rbrace\ .
\end{equation} 

The two latter relations are the independent conditions to close both the
radiative transfer and the spline chain.

\parindent=3em
According to the cubic approximation for $S(\tau)$, 
$S^{\prime}(\tau_{NL})$ is a linear function of $S(\tau_{NL-1})$, 
$S(\tau_{NL})$, $S^{\prime\prime}(\tau_{NL-1})$ and
$S^{\prime\prime}(\tau_{NL})$, as shown by eq. (9). The ''physical'' equation
(45) and the spline equation (9) lead to a new linear relation among
$S(\tau_{NL-1})$, $S^{\prime\prime}(\tau_{NL-1})$,
$S(\tau_{NL})$ and $S^{\prime\prime}(\tau_{NL})$ that, together with eq.s (39),
(40) and (44) lead easily to the explicit values of the latter four variables.
Consequently we easily obtain also the values of $S^{\prime}(\tau_{NL})$ and
$S^{\prime\prime\prime}(\tau_{NL})$. The explicit values of these variables at
$\tau_{NL}$ allow us to compute those of the set 
$\lbrace I^{+}(\tau_{NL},\mu_{J})\rbrace$ through eq. (5).

Once the explicit values of $S(\tau_{NL-1})$, $S(\tau_{NL})$, 
$S^{\prime\prime}(\tau_{NL-1})$, $S^{\prime\prime}(\tau_{NL})$ as well as those
of the set $\lbrace I^{+}(\tau_{NL},\mu_{J}, J=1,ND)\rbrace$ are known, it is 
straightforward to compute those of the set
$\lbrace I^{+}(\tau_{NL-1}),\mu_{J}, J=1,ND)\rbrace$ via eq. (15). 
Then eq.s (27) and (28) will yield the explicit values of $S(\tau_{NL-2})$ and 
$S^{\prime\prime}(\tau_{NL-2})$, hence those of the set
$\lbrace I^{+}(\tau_{NL-2},\mu_{J}, J=1,ND)\rbrace$. And so on along the back-
substitution. 

\section{Conclusions}

Our Implicit Integral Method is based on the progressive treatment of the 
different layers that consitute a model of the stellar atmosphere physical
system, from the outermost layer (the surface) to the deepest one (the
bottom). The protagonist variables of the method are the upgoing and downgoing
specific intensities $I^{\pm}(\tau,\mu)$ as well as the corresponding source
functions that besides the thermal sources include a scattering-like integral
into which there
enter the foregoing specific intensities. Precisely, the study (and
the elimination) of each single layer $(\tau_{L},\tau_{L+1})$ leads to a
relation that links linearly the value $S(\tau_{L})$ of the source function at
$\tau_{L}$ with $S(\tau_{L+1})$, the value at $\tau_{L+1}$. Once obtained via 
the study of the last layer the
relation between the values of the source function at the two last optical depth
points, the boundary condition  at $\tau_{NL}$ given by eq. (5) makes it
possible to compute the explicit values of $S(\tau_{NL})$ and $S(\tau_{NL-1})$,
hence all the others.

In order to design the required elimination scheme it is necessary to employ a
mathematical model for $S(\tau)$. In principle the simplest and easiest model
would be a piece-wise linear one, but the discontinuity of the first
derivative $S^{\prime}(\tau)$ at each knot $\tau_{L}$ can imply severe errors
and possible numerical instabilities because the above discontinuity is 
incompatible with the radiative transfer (RT) process itself, where both 
$I^{\pm}(\tau,\mu)$ and their first derivatives must be continuous, and
therefore also the mean intensity $J(\tau)$ and its first derivative. Thus
the foregoing model cannot be correct,
but for extreme cases of the thermal sources.

A piece-wise parabolic model warrants the continuity of $S^{\prime}(\tau)$ at
all depth points. Such a model shall include also $S^{\prime}(\tau)$ as a
protagonist variable in the process of progressive elimination of the
atmospheric layers. Hence $S^{\prime}(\tau)$ must be put into relation with
the foregoing protagonist variables, which can be done either mathematically or
physically.

From the mathematical standpoint we can introduce $S^{\prime}(\tau)$ by means
of the formula
\begin{equation}
S^{\prime}(\tau_{L})\ =\ 2\ \frac{S(\tau_{L+1}) - S(\tau_{L})}
{\tau_{L+1} - \tau_{L}}\ -\ S^{\prime}(\tau_{L+1})\ ,
\end{equation}

\parindent=0em
which could however introduce numerical instabilities because of the difference
between the two terms in the right-hand side, above all in the back-substitution
process that works with explicit values. On the other hand,
from the physical standpoint $S^{\prime}(\tau)$ could be included by taking
into account at all the optical depth points the equations (12) and (13) for the
derivatives of the source function. However, in
case that $\varepsilon(\tau)$ and $B(\tau)$ show large variations (as it is the
case of the formation of Lyman $\alpha$ in cool stars), severe instabilities may
appear, too. 

\parindent=3em
These drawbacks can be avoided by introducing a piece-wise cubic approximation,
where a further
protagonis variable has to be included, namely the second derivative
$S^{\prime\prime}(\tau)$. That is, by means of a cubic spline model that 
automatically warrant the continuity of $S(\tau)$ and its first two 
derivatives. To circumvent the explicit calculation of the derivatives makes the
above difficulties vanish.

The source function $S(\tau_{L})$ at each depth point $\tau_{L}$ will be
expressed as a linear function of $S(\tau_{L+1})$ and 
$S^{\prime\prime}(\tau_{L+1})$. Therefore we shall transmit from any optical
depth to the next one also the (implicit) value of 
$S^{\prime\prime}(\tau_{L+1})$. This is achieved thanks to the fundamental
relation that assures the continuity properties imposed by the cubic spline
condition (cf. eq. [8]). However the propagation of $S(\tau)$ and
$S^{\prime\prime}(\tau)$ via a cubic spline model constitutes a two-point
boundary problem. Nevertheless this is perfectly compatible with the treatment 
of the
transmission of the specific intensities, as it is performed in the scheme for
the solution of the two-point boundary value RT problem. Both propagation
processes can be treated simultaneously. 

Under these conditions we can warrant the elimination of many of the causes
of instability that can spoil the algorithm for the solution of the system of
specific RT equations coupled through a scattering-like term in the source
function, whose initial conditions are assigned at different points of the
physical system.

\end{document}